# ON THE THEORY OF TURBULENT FLAME VELOCITY

Vitaly Bychkov[1], V'yacheslav Akkerman[1,2], and Arkady Petchenko[1]

[1]Institute of Physics, Umeå University, SE-901 87 Umeå, Sweden

[2]Nuclear Safety Institute (IBRAE) of Russian Academy of Sciences,
B. Tulskaya 52, 115191 Moscow, Russia

**Abstract**

The renormalization ideas of self-similar dynamics of a strongly turbulent flame front are applied to the case of a flame with realistically large thermal expansion of the burning matter. In that case a flame front is corrugated both by external turbulence and the intrinsic flame instability. The analytical formulas for the velocity of flame propagation are obtained. It is demonstrated that the flame instability is of principal importance when the integral turbulent length scale is much larger than the cut off wavelength of the instability. The developed theory is used to analyse recent experiments on turbulent flames propagating in tubes. It is demonstrated that most of the flame velocity increase measured experimentally is provided by the large scale effects like the flame instability, and not by the small-scale external turbulence.

**Introduction**

One of the most important problems of combustion science is the problem of the velocity of turbulent premixed flames. The question is how the turbulent flame speed (or burning rate) $U_w$ depends on the intensity of external flow characterized by the rms-velocity in one direction $U_{rms}$ [1-4]. Typically, the turbulent flame speed $U_w$ is much larger than the planar flame velocity $U_f$, which is determined by thermal and chemical parameters of the burning mixture. Particularly, in the traditional model of an infinitely thin flame front propagating with a constant normal velocity $U_f$, the ratio $U_w/U_f$ is equal to the relative increase of the flame surface area because of the corrugated flame shape. Such an approach is possible in the flamelet regime of turbulent burning, when a flame front is strongly corrugated on large length scales in comparison with the flame thickness $L_f$, but it retains its internal structure similar to the laminar one. For a long time, since the works by Damkohler and Shelkin, it was a common belief that the scaled turbulent flame speed $U_w/U_f$ may be expressed as a function of the scaled rms turbulent velocity $U_{rms}/U_f$ only

$$U_w/U_f = F(U_{rms}/U_f) \qquad (1)$$

independent of other dimensionless parameters of the flow [1]. Following the assumption (1) researchers tried to obtain the function $F$ experimentally or in the simple theoretical models [1, 6]. The absolute majority of theoretical works devoted to the turbulent flame velocity was performed in the artificial model of zero thermal expansion at the flame front, when density of the fuel mixture $\rho_f$ is the same as the density of the burning products $\rho_b$, with the expansion factor $\Theta \equiv \rho_f/\rho_b = 1$ [6-12]. Such a model is quite far from realistic laboratory flames, for which the density of burning matter drops almost by an order of magnitude $\Theta = 5-8$. One more consequence of the assumption (1) is that the turbulent flame speed should be independent of a particular geometry of experimental studies. In that case the experimental results for $U_w/U_f$ may be compared directly for different experiments

3without any preliminary analysis, nor matter if the measurements have been performed in a combustion bomb [5], for Bunsen flames [13-15], for plate-stabilized burning [16], for burning in tubes [17, 18] or in other configurations. However, it was demonstrated already in the review [5] that the simple assumption (1) cannot describe the diversity of measurements of turbulent flame speed. Instead of a single curve (1), experimental points of $U_w/U_f$ versus $U_{rms}/U_f$ obtained by different groups looked like a wide cloud. The work [5] tried to organize the cloud by introducing empirically additional parameters into the dependence (1) such as the Karlovitz number and/or the Reynolds number of the flow. On the other hand, recent theoretical results on the turbulent flame velocity for realistically large thermal expansion have shown that $U_w/U_f$ depends on a large number of parameters like the expansion factor $\Theta$, the Markstein number $Mk$, the turbulent spectrum, the integral turbulent length scale, the maximal hydrodynamic length scale of the flow, and many others [19-21]. This diversity of parameters came to play because of several different physical mechanisms, which influence the flame dynamics in the case of realistically large thermal expansion. Among these mechanisms we should mention, first of all, the hydrodynamic Darrieus-Landau (DL) instability [1, 22] and fast burning along the vortex axis [23]. For example, the experiments [13-15] demonstrated that at certain conditions the turbulent flame velocity may increase noticeably because of the DL instability. According to these measurements, even in the case of zero turbulence, the velocity of flame propagation $U_w$ may be 6 times larger than the planar flame velocity $U_f$ due to the DL instability only. The theory [19-21] indicated also the important role of the DL instability in many experimental configurations. However, the most important conclusion, which follows both from the experiments and from the recent theory, is that the values $U_w/U_f$ measured in different experiments depend on a particular experimental configuration and must be analyzed separately. Indeed, in different experimental flows the relative roles of the external turbulence, of the DL instability, and other processes are different, which requires special approach to any configuration. For example, it was demonstrated in [19] that in the case of spherical flames expanding from the center of a combustion bomb, the DL instability and external turbulence work together with their relative strength depending on the Karlovitz number. The theoretical analysis [19] explained the empirical dependence of the turbulent flame speed on the Karlovitz number proposed in [5].

In this paper we present our recent results on the theory of turbulent flame velocity with realistic density drop at the front. We use the theoretical results to analyse the experimental data [18] on turbulent burning in tubes. We demonstrate that most of the flame velocity increase in [18] is provided by the large-scale effects: by the DL instability and by non-slip at the walls. Only rather moderate part of the flame velocity increase happens because of the small-scale external turbulence, which is opposite to the general belief.

**Scale-invariance for a turbulent "flame" with zero thermal expansion**

According to the well-known Clavin-Williams formula [6], obtained in the case of no thermal expansion $\Theta = 1$ and zero flame thickness, weak turbulence increases the velocity of a turbulent flame as $\Delta U = U_w - U_f$ with

$$\Delta U / U_f = U_{rms}^2 / U_f^2, \qquad (2)$$

where $U_{rms} = u_{rms}(\lambda_T)$ and $\lambda_T$ is the integral length scale of the turbulent flow. Equation (2) may be also presented with the help of spectral density $\varepsilon_T(k)$ of the turbulent kinetic energy

$$\Delta U / U_f = U_f^{-2} \int_{k_T}^{k_\nu} \varepsilon_T(k) \, dk, \qquad (3)$$

where

$$u_{rms}^2 = \int_{k}^{k_\nu} \varepsilon_T(k) \, dk, \qquad U_{rms}^2 = \int_{k_T}^{k_\nu} \varepsilon_T(k) \, dk, \qquad (4)$$

$k_T = 2\pi / \lambda_T$ and $k_\nu = 2\pi / \lambda_\nu$ are the wave numbers corresponding to the integral and Kolmogorov (dissipation) length scales, $\lambda_T$ and $\lambda_\nu$, respectively; the former being usually much larger than the latter $\lambda_T \gg \lambda_\nu$, $k_T \ll k_\nu$. In the case of the Kolmogorov turbulent spectrum $\varepsilon_T(k) \propto k^{-5/3}$ we find from Eq. (4)

$$\varepsilon_T(k) = (2/3) U_{rms}^2 k_T^{2/3} k^{-5/3}. \tag{5}$$

Equation (2) has been extrapolated to the case of a strongly turbulent flame assuming self-similar properties of the corrugated front [8]. Following [8] we decompose the turbulent flame wrinkles into components with different wave numbers (narrow bands in the spectrum), each of them providing similar small increase of the flame front velocity. Let us designate the velocity of flame propagation corresponding to the wrinkles with the wave numbers above $k$ by $U = U(k)$. Since every band in the spectrum of flame wrinkles leads to infinitesimal increase in the flame velocity, then we can write the velocity increase for any band in the form similar to Eq. (2)

$$dU/U = -\varepsilon_T(k) dk / U^2. \tag{6}$$

Integrating Eq. (6) over the whole turbulent spectrum one obtains the propagation velocity $U_w$ of a strongly corrugated flame with zero thermal expansion $\Theta = 1$ [8]

$$U_w^2 = U_f^2 + 2 U_{rms}^2. \tag{7}$$

**Strongly corrugated flames produced by the DL instability only**

Let us consider similar scale-invariant formulas for the case of a flame front corrugated because of the DL instability only, when there is no external turbulence. It has been obtained experimentally [24, 25] that a corrugated spherical flame front, unstable according to the DL mechanism, accelerates with the velocity of flame propagation $U_w$ depending on the characteristic length scale of the hydrodynamic motion $\lambda$ as

$$U_w = U_f (\lambda / \lambda_c)^D, \tag{8}$$

where $\lambda_c$ is the cut off wavelength of the DL instability proportional to the flame thickness, and $D$ is a constant power exponent. The self-similar flame acceleration has been interpreted as development of a fractal structure at the flame front with wrinkles of different wave numbers imposed on each other [24, 25]. The theoretical studies of the stability properties of curved flames [22, 26] lead to similar conclusions. In that case the cut off wavelength of the DL instability $\lambda_c$ plays the role of the inner cut off in the fractal cascade, and $D$ is the excess of the fractal dimension of the flame front over the embedding dimension (the embedding dimension is obviously 2 for the realistic experiments with three-dimensional flows). According to the experimental measurements [24, 25], the fractal excess is approximately $D \approx 1/3$ for all investigated laboratory flames. The theoretical estimates [22, 26] suggest that the fractal excess depends on the expansion factor $D = D(\Theta)$ with $D \approx 1/3$ for $\Theta = 5-8$, typical for laboratory flames, and $D \to 0$ when $\Theta \to 1$. Assuming self-similar properties of the fractal cascade, we should expect that the "intermediate" velocity of flame propagation $U = U(k)$ produced by the wrinkles with wave numbers in between $k$ and $k_c = 2\pi/\lambda_c$ depends on $k$ as

$$U = U_f (k/k_c)^{-D}. \tag{9}$$

The last equation may be also presented in the differential form similar to Eq. (7)

$$dU/U = -\varepsilon_{DL}(k)\, dk, \tag{10}$$

with $\varepsilon_{DL} = D/k$ for $k < k_c$ and $\varepsilon_{DL} = 0$ when $k \geq k_c$.

**The external turbulence and the DL instability work together**

In general, if a flame front with realistic thermal expansion $\Theta = 5-8$ propagates in a turbulent flow, then both the DL instability and external turbulence contribute to the velocity increase. Thus, the velocity increase depends both on the scaled turbulent intensity $U_{rms}^2/U_f^2$ and on the intrinsic parameters of the flame dynamics such as $\Theta$:

$$\Delta U/U_f = F(U_{rms}^2/U_f^2; \Theta; ...). \tag{11}$$

If the turbulence is weak $U_{rms}^2/U_f^2 \ll 1$, then using Taylor expansion we reduce Eq. (11) to

$$\Delta U / U_f = C_{DL}(\Theta;...) + C_T(\Theta;...) U_{rms}^2 / U_f^2. \tag{12}$$

Obviously, the first term in Eq. (12) describes the velocity increase due to the DL instability only $C_{DL}(\Theta;...) = (\Delta U / U_f)_{DL}$. The second term determines contribution by external turbulence, with the coefficient $C_T$ in Eq. (12) determined by two different physical processes. First, the flame front is drifted and distorted by vortices perpendicular to the direction of flame propagation; this is a kinematical effect designated in the following by the label "$\perp$". Second, vortices parallel to the direction of flame propagation also increase the flame velocity. In that case the centrifugal acceleration created by a vortex acts like an effective gravitational field and makes a flame front curved; we designate this effect by "$\parallel$". Respectively, the coefficient $C_T$ consists of two parts

$$C_T = C_\perp + C_\parallel. \tag{13}$$

The coefficient $C_\perp$ has been obtained analytically in [20] as

$$C_\perp = 4\Theta^2 \frac{[1 + L_f k(C_3 + Mk)]^2 + [1 + L_f k(C_3 - Mk)]^2}{[(\Theta+1)(1 + L_f C_1 k) + \Theta(\Theta-1)(1 - \lambda_c k / 2\pi)]^2 + 4\Theta^2 (1 + L_f C_2 k)^2}, \tag{14}$$

where $Mk$ is the Markstein number. The cut-off wavelength of the DL instability $\lambda_c$ and the coefficients $C_1$, $C_2$, $C_3$ have been calculated in the linear theory of flame response to weak external turbulence [27] as

$$C_1 = \frac{\Theta-1}{\Theta+1} Mk - \frac{\Theta}{\Theta+1} J_1, \quad C_2 = \Theta(Mk - J_1), \tag{15}$$

$$C_3 = (\Theta-1)Mk - \Theta J_1 - \Pr(h_b - 1), \tag{16}$$

$$\lambda_c = 2\pi L_f \left[ h_b + \frac{3\Theta-1}{\Theta-1} Mk - \frac{2\Theta}{\Theta-1} J_1 + (2\Pr-1)\left(h_b - \frac{J_2}{\Theta-1}\right) \right], \tag{17}$$

where

$$J_1 = \int_1^\Theta \vartheta^{-1} h(\vartheta) d\vartheta, \qquad J_2 = \int_1^\Theta h(\vartheta) d\vartheta. \tag{18}$$

Following [27] we describe the temperature dependence of the thermal conduction coefficient by using the function $h(\vartheta) = \sqrt{\vartheta}$, with $h(1) = 1$ and $h(\Theta) = h_b = \sqrt{\Theta}$, where $\vartheta = T/T_f$ is the scaled temperature of the burning matter $1 \leq \vartheta \leq \Theta$. Formulas (16), (17) include also the Prandtl number Pr characterizing the relative role of viscosity and thermal conduction. In the limit of negligible flame thickness $kL_f \ll 1$, $k\lambda_c \ll 1$ the coefficient $C_\perp$ is described by the simple formula

$$C_\perp = \frac{8\Theta^2}{(\Theta^2 + 1)^2 + 4\Theta^2}. \tag{19}$$

The coefficient $C_\parallel$ has been found in a semi-analytical way in [28] as

$$C_\parallel = C_{\parallel,\infty}(1 - k/k_c). \tag{20}$$

The value $C_{\parallel,\infty}$ depends on thermal expansion $\Theta$; for example, for $\Theta \approx 7.5$ it was calculated numerically as $C_{\parallel,\infty} \approx 0.5$. The coefficients $C_\perp$ and $C_\parallel$ are shown in Fig. 1 versus the expansion factor $\Theta$. As we can see from Fig. 1, in the case of realistic thermal expansion $\Theta = 5 - 8$ and infinitely thin flame front the coefficient $C_\perp$ is rather small $C_\perp = 0.1 - 0.15$, which is about $7 - 10$ times smaller than the value $C_\perp = 1$ calculated for $\Theta = 1$ according to the Clavin-Williams formula, Eq. (2). Thus, the role of the perpendicular vortices is almost an order of magnitude weaker than it was believed previously. On the contrary, the role of parallel vortices is noticeably stronger, with $C_{\parallel,\infty} \approx 0.5$. When both effects work together, the coefficient $C_T$ in Eq. (12) becomes comparable to unity. Besides, both coefficients $C_\perp$ and $C_\parallel$ depend on the characteristic length scale of the flow $\lambda$. For example, Fig. 2 presents $C_\perp$ and $C_\parallel$ versus the flow length scale $\lambda$ for methane-air flames. An important property of both values $C_\perp$ and $C_\parallel$ is that these coefficients go rather fast to zero at the length scales below the DL cut-off $\lambda < \lambda_c$, and they tend to some saturation values $C_{\parallel,\infty}$ and $C_{\perp,\infty}$ for $\lambda \gg \lambda_c$. The condition $\lambda \gg \lambda_c$ is the only rigorous limit for the renormalization analysis. In that case one has to take $C_T$ in the form of a Heviside step-function with $C_T = 0$ for $\lambda < \lambda_c$ and $C_T = C_{T,\infty}$ for $\lambda > \lambda_c$. Taking into account Eq. (12) and assuming scale-invariance of the flame dynamics, we find the increase of the turbulent flame velocity produced by one narrow band in the turbulent spectrum

$$dU/U = -\varepsilon_{DL}(k)\,dk - C_T \varepsilon_T(k)\,dk/U^2, \tag{21}$$

or

$$d(U^2)/dk = -2\varepsilon_{DL}(k)U^2 - 2C_T \varepsilon_T(k). \tag{22}$$

Thus we came to a linear differential equation with coefficients depending on the variable, Eq. (22), which may be solved analytically by a standard method. Below we will consider the solution to Eq. (22) for the most typical case $\lambda_m > \lambda_T > \lambda_c > \lambda_\nu$, when the maximal length scale of the flow is larger than the turbulent integral length scale, and the DL cut-off is larger than the Kolmogorov cut-off. We also chose the excess of the fractal flame dimension $D = 1/3$. A more general solution to Eq. (22) is presented in [19].

To find the solution to Eq. (22) we have to consider separately three domains of large, moderate and small wave numbers, $k_c < k < k_\nu$, $k_T < k < k_c$ and $k_m < k < k_T$, respectively. Adopting $C_T$ in the form of a Heviside function, we obtain a planar flame front in the domain $k_c < k < k_\nu$, since the DL instability does not develop at such a short wavelengths and influence of external turbulence is effectively suppressed by thermal conduction. Thus, for $k_c < k$ we obtain $U(k) = U_f$. Taking into account this result, the solution to Eq. (22) in the domain $k_T < k < k_c$ may be written as

$$U^2 = U_f^2 (k/k_c)^{-2D} + 2C_T\, k^{-2D} \int_k^{k_c} \eta^{2D} \varepsilon_T(\eta)\,d\eta. \tag{23}$$

In the case of Kolmogorov turbulence $\varepsilon_T(k) \propto k^{-5/3}$ and $D = 1/3$ we find from Eq. (23)

$$U^2 = U_f^2 (k/k_c)^{-2/3} + (4/3) C_T U_{rms}^2 (k/k_T)^{-2/3} \ln(k_c/k). \tag{24}$$

Taking into account the whole spectrum of turbulent pulsations with the low integration limit $k = k_T$ in Eq. (24) we find the velocity of flame propagation for $\lambda = \lambda_T$

$$U^2 = U_f^2 (\lambda_T / \lambda_c)^{2/3} + (4/3) C_T U_{rms}^2 \ln(\lambda_T / \lambda_c). \tag{25}$$

Finally, we have to perform integration in the limit of large wavelengths $\lambda_m > \lambda > \lambda_T$ (small wave numbers $k_m < k < k_T$). In that case we have $\varepsilon_T(k) = 0$, and, from the renormalization point of view, the formula (25) plays the role of an effective "planar flame velocity". Integrating Eq. (22) we come to the formula similar to Eq. (8)

$$U_w^2 = \left[ U_f^2 (\lambda_T / \lambda_c)^{2/3} + (4/3) C_T U_{rms}^2 \ln(\lambda_T / \lambda_c) \right] (\lambda_m / \lambda_T)^{2/3}, \tag{26}$$

or

$$U_w^2 = U_f^2 (\lambda_m / \lambda_c)^{2/3} + (4/3) C_T U_{rms}^2 (\lambda_m / \lambda_T)^{2/3} \ln(\lambda_T / \lambda_c). \tag{27}$$

When turbulent intensity is zero $U_{rms} = 0$, then Eq. (27) goes over to the velocity increase produced by the DL instability only, Eq. (8). Comparing the second terms in the velocity increase in Eq. (7) and Eq. (27) (the terms related to the external turbulence) we can see that the turbulent term in Eq. (27) is multiplied now by a large factor $(\lambda_m / \lambda_T)^{2/3}$, which makes the influence of external turbulence much stronger in presence of the strong DL instability. It follows from Eq. (27) that the tangent of the plot $U_w$ versus $U_{rms}$ studied in numerous experiments, is not a universal property of a turbulent flow. On the contrary, this tangent is determined by the strength of the DL instability and other large-scale effects, which typically differ from one experiment to another, or even from one experimental point to another for the same installation. In that sense the obtained result, Eq. (26), is qualitatively different from the universal law Eq. (1), and it explains the diversity of experimental data measured by different research groups.

**Comparison to the experiments**

As it was shown above, every experiment on turbulent flame velocity requires separate approach and separate analysis. As an illustration, below we consider the experiments [18], which investigated flame propagation in tubes similar to the present theory. In these experiments the propane-air flames with the equivalence ratio $\phi = 0.75; 1$ propagated in a tube with a rectangular cross-section $9\,cm \times 3.5\,cm$. According to [18] the integral turbulent length of the flow $\lambda_T \approx 0.5\,cm$ was much smaller than the tube width. Using the data for the Markstein number from [29] we calculated the DL cut-off wavelength as $\lambda_c = 0.21\,cm$ for $\phi = 1$ and $\lambda_c = 0.49\,cm$ for $\phi = 0.75$. As we can see, in these experiments the DL cut-off is comparable to the turbulent integral length scale, which is rather typical for the experiments on turbulent flame velocity. In that case, of course, the renormalization analysis cannot be applied rigorously, but it can be used only formally. Besides, in that case we cannot take the coefficients $C_\perp$ and $C_\parallel$ in the form of a Heviside step-function. Instead, we have to take into account dependence of these coefficients on the length scale $\lambda$, Eqs. (14) and (20), and to integrate Eq. (22) numerically. The result of numerical integration is compared in Fig. 3 to the experimental data for $\phi = 1$. To understand the results better, we have performed integration first for small length scales up to $\lambda_T = 0.5\,cm$, and, second, for all length scales up to $\lambda_m = 2 \times 9\,cm$. As we can see from Fig. 3, the flame velocity increase provided by wrinkles of small length scales (that is, by external turbulence) is well below the experimental points. However, even in that velocity increase, a good deal is provided by the DL instability of small scales, since the plot starts at $U_w \approx 1.7 U_f$ for $U_{rms} = 0$. It demonstrates that the role of external turbulence is rather moderate in the experimental measurements of the turbulent flame velocity. On the contrary, performing integration for all length scales we come much closer to the experimental points. The additional velocity increase is determined by the large-scale DL instability similar to the factor obtained in Eq. (26) in comparison with Eq. (25). Still, even these theoretical results are somewhat below the experimental ones. To complete the comparison, we have also taken into account influence of the non-slip boundary conditions at the walls. This effect has been investigated in our recent paper [30]; it provides additional increase of the flame velocity by a factor of about 1.5. As we can see, taking all effects of the large scale into account we obtain very good agreement of the theory with the experiments. Figure 4 provides the final comparison of the theory and the experiments for both values of the equivalence ratio $\phi = 0.75; 1$ used in experiments [18]: in both cases the agreement is quite good.

## Summary

We have developed the ideas of the self-similar behaviour of a strongly corrugated flame front to the case of a flame influenced both by the external turbulence and by the DL instability. The obtained analytical formula Eq. (27) demonstrates that the DL instability is of principal importance when the integral turbulence length scale is large. The developed theory is used to analyse recent experiments on turbulent flames propagating in tubes. It is demonstrated that most of the flame velocity increase measured experimentally is provided by the large scale effects like the flame instability, and not by the small-scale external turbulence.

## Acknowledgements

This work was supported by the Swedish Research Council (VR) and by the Kempe Foundation.

**Figure captions**

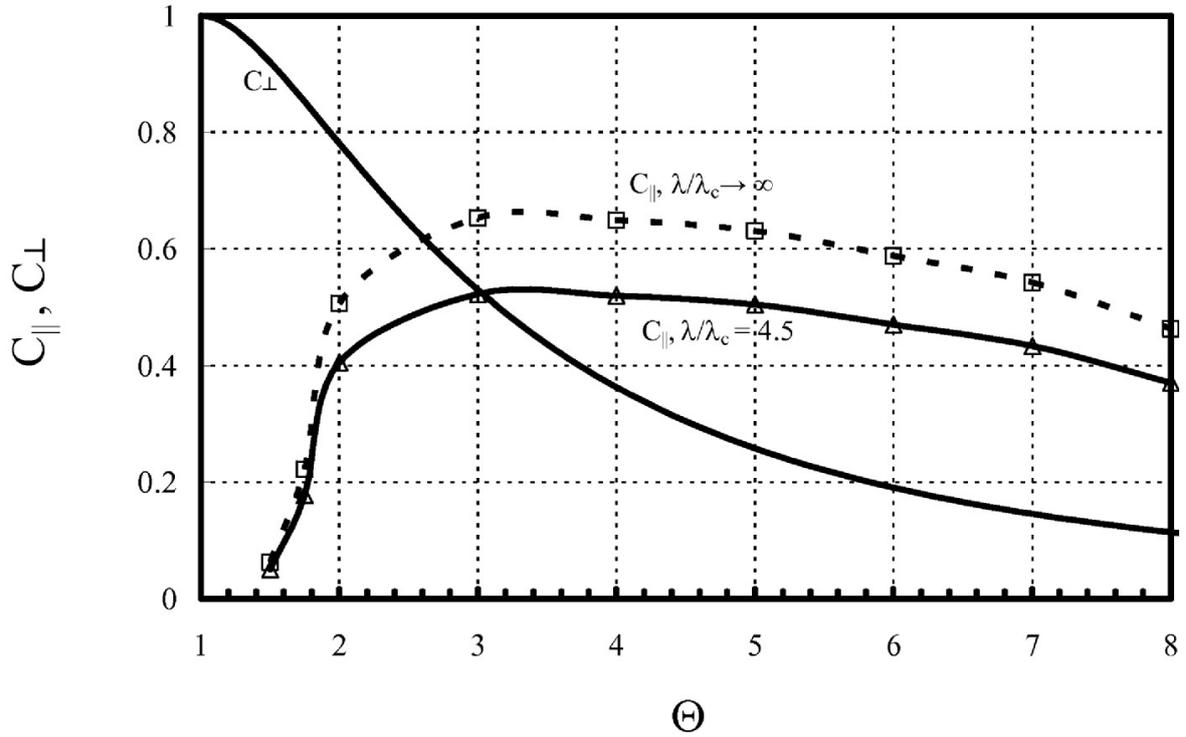

Fig. 1    BYCHKOV ET AL        On the theory of turbulent flame velocity

**Figure 1**. The coefficients $C_\perp$ and $C_\parallel$, Eqs. (13) and (14), versus the expansion factor $\Theta$. The dashed line presents the value $C_{\parallel,\infty}$.

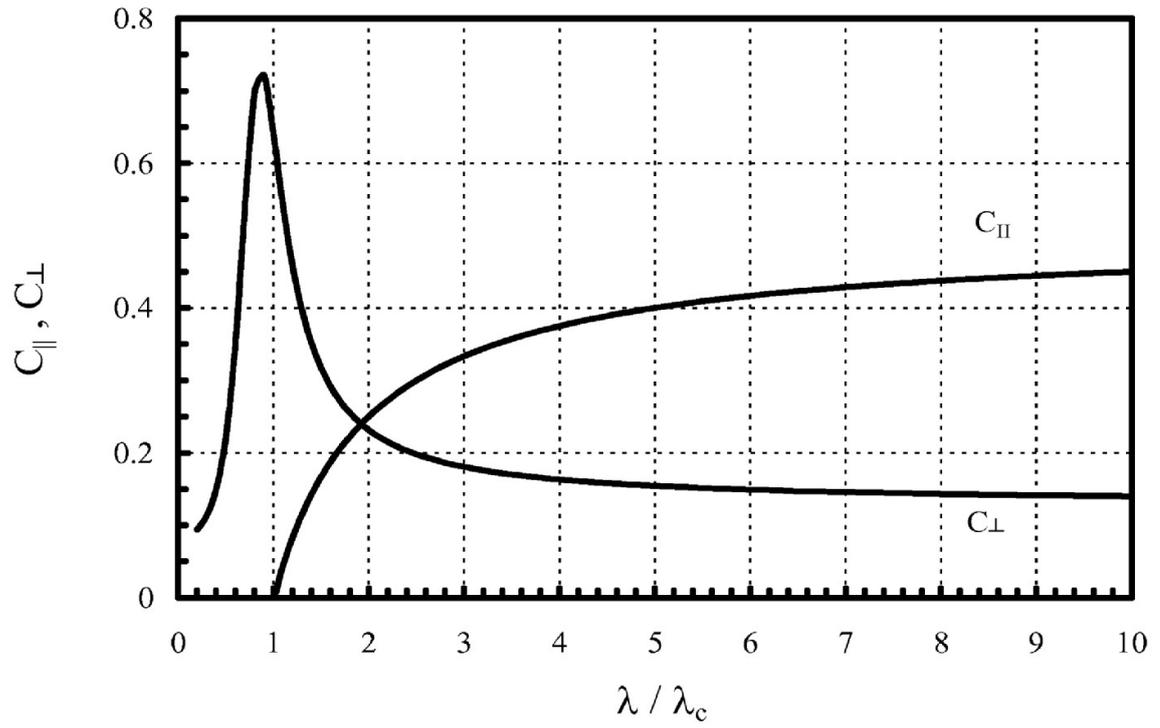

Fig. 2    BYCHKOV ET AL    On the theory of turbulent flame velocity

**Figure 2**. The coefficients $C_\perp$ and $C_\|$, Eqs. (13) and (14), versus the scaled flow length scale $\lambda/\lambda_c$ for the methane-air flames with the expansion factor $\Theta = 7.48$, the Markstein and Prandtl numbers $Mk = 3.73$, $Pr = 0.7$.

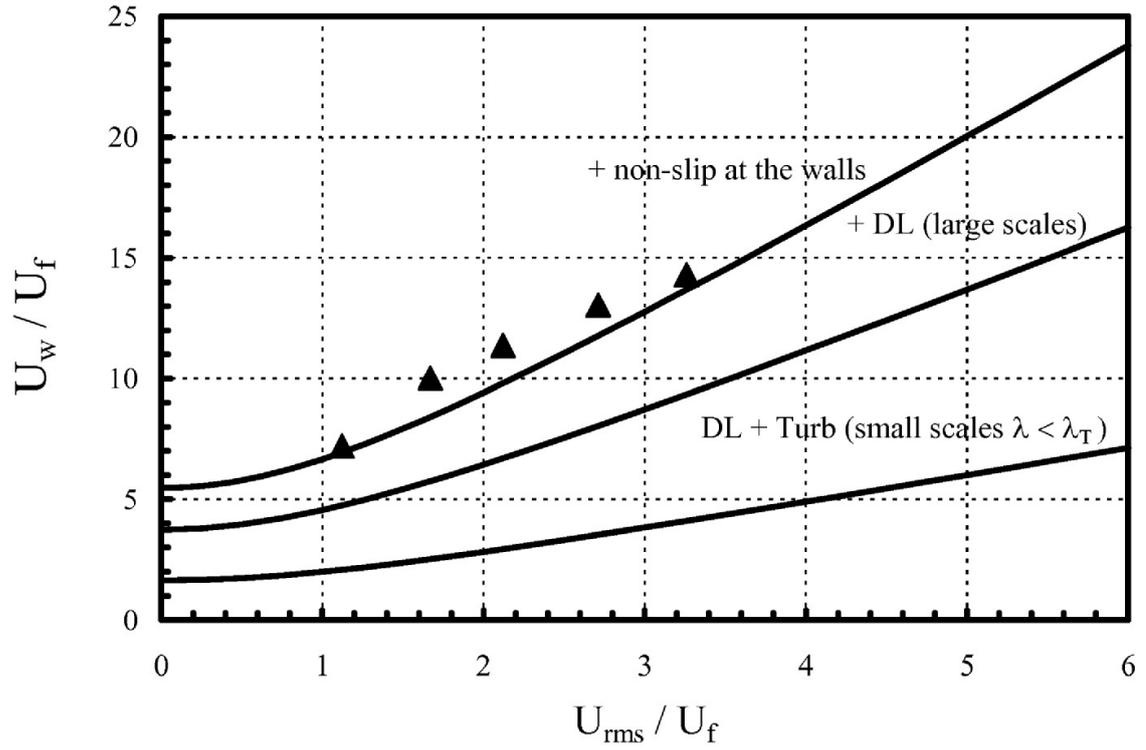

Fig. 3    BYCHKOV ET AL    On the theory of turbulent flame velocity

**Figure 3**. Comparison of the theory (solid lines) to the experiments [18] with stoichiometric propane-air flames.

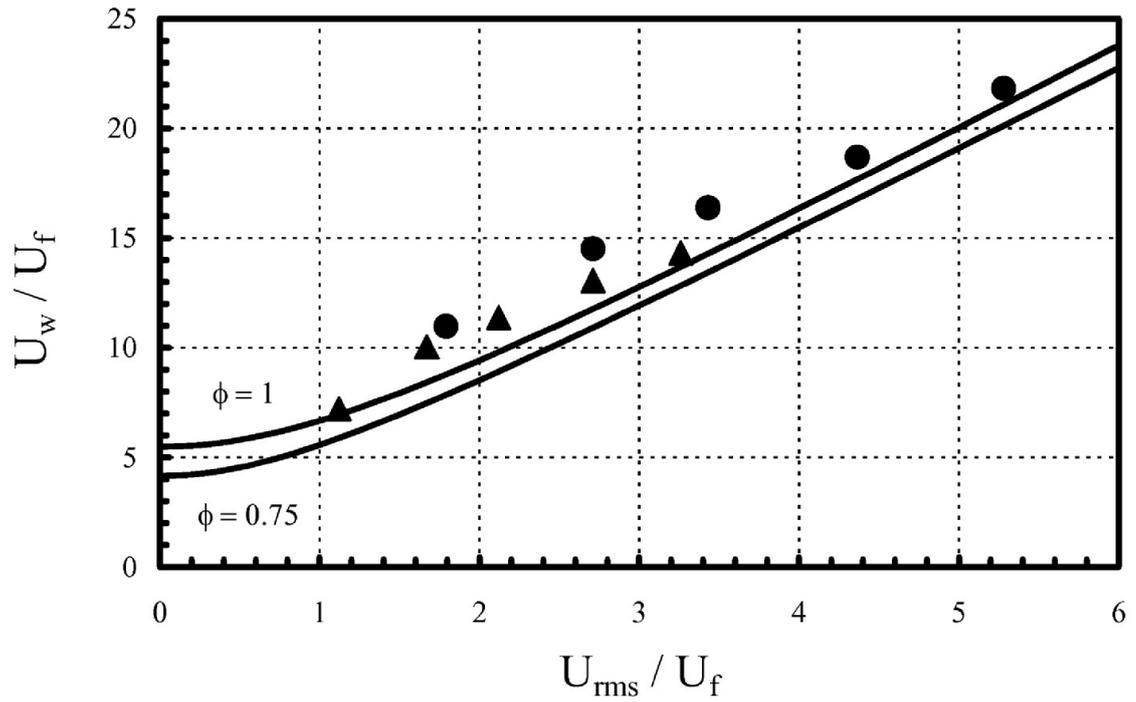

Fig. 4  BYCHKOV ET AL    On the theory of turbulent flame velocity

**Figure 4**. Comparison of the theory (solid lines) to the experiments [18] with stoichiometric (triangles) and non-stoichiometric (circles) propane-air flames.